# ON A $p$-LAPLACIAN TYPE OF EVOLUTION SYSTEM AND APPLICATIONS TO THE BEAN MODEL IN THE TYPE-II SUPERCONDUCTIVITY THEORY

## HONG-MING YIN

ABSTRACT. In this paper we study the Cauchy problem for an $p$−Laplacian type of evolution system $\mathbf{H}_t + \nabla \times [|\nabla \times \mathbf{H}|^{p-2}\nabla \times \mathbf{H}|] = \mathbf{F}$. This system governs the evolution of a magnetic field $\mathbf{H}$, where the current displacement is neglected and the electrical resistivity is assumed to be some power of the current density. The existence, uniqueness and regularity of solutions to the system are established. Furthermore, it is shown that the limit solution as the power $p \to \infty$ solves the problem of Bean's model in the type-II superconductivity theory. The result provides us information about how the superconductor material under the external force to become the normal conductor and vice visa. It also provides an effective method to find numerical solutions to Bean's model.

## 1. INTRODUCTION

In this paper we shall study the following degenerate evolution system:

$$\mathbf{H}_t + \nabla \times [|\nabla \times \mathbf{H}|^{p-2}\nabla \times \mathbf{H}] = \mathbf{F}, \qquad (x,t) \in Q_T, \qquad (1.1)$$

$$\nabla \cdot \mathbf{H} = 0, \qquad (x,t) \in Q_T, \qquad (1.2)$$

$$\mathbf{H}(x,0) = \mathbf{H}_0(x), \qquad x \in R^3, \qquad (1.3)$$

where $Q_T = R^3 \times (0,T]$ for some $T > 0$ and $p > 2$. Here a bold letter represents a vector in $R^3$.

The system (1.1) is derived from Maxwell's system where the current displacement is neglected since it is small in comparison of eddy currents ( see *Landau-Lifschitz* [12]). The vector $\mathbf{H}$ represents the magnetic field. The electrical resistivity is assumed to be equal to some power of the current density. The motivation for our investigation of (1.1)-(1.3) is twofold. On one hand, the system (1.1) is a natural generalization of the scalar p-Laplacian which has been studied by many authors (see *DiBenedetto* [6] and the references therein). More recently, the limit solution of the p-Laplacian equation as $p \to \infty$ is used to describe the fast-slow diffusion model and the sand-pile collapsing phenomenon (see *Arronson-Evans-Wu* [1], *Evans-Gangbo* [9], *Evans-Feldman-Gariepy* [8] and the references therein). On the other hand, it will be seen that the limit solution to the system (1.1)-(1.3) solves the problem of Bean's critical-state model for type-II superconductor (see *Bean* [3]). In Bean's model, the current density is assumed to be always less than or equal to a critical current, denoted by $J_c$. The superconductive region is characterized by the current density $J = |\nabla \times \mathbf{H}|$. When the current density is strictly less than the critical number $J_c$, then there is

Research at MSRI is supported in part by NSF grant DMS-9701755.

no resist for the movement of electrons (the superconductivity is achieved) while the normal conductor region is the one where $|J| = J_c$. When a magnetic field is under the influence of an external force, a normal conductor material will become a superconductor one or vice visa. The motion of the interface is driven by the external force. Recently, Prigozhin in [14] reformulated the problem to a variational inequality and established the existence of a unique weak solution to the variational inequality if the space dimension is equal to two. Some numerical issues are discussed in [2] (also see a more recent paper [15]). Surprisingly, we show that the weak solution defined in [14] can be obtained as the limit of the solution to the system (1.1) as $p \to +\infty$. Here we assume that the Bean's critical number $J_c = 1$ without loss of generality. Our argument is entirely different from that of [14]. From the physical point of view, for large $p$ the electric resistivity is small in the region $S_\varepsilon = \{(x, t) : |\nabla \times \mathbf{H}| \leq 1 - \varepsilon\}$ while it is very large in $\{(x, t) : |\nabla \times \mathbf{H}| \geq 1 + \varepsilon\}$. Thus, the resistivity in $S_\varepsilon$ becomes smaller and smaller as $p$ increases. And $S_\varepsilon$ becomes the superconductor region as $\varepsilon \to 0$ (no resistivity). The region $\{(x, t) : 1 - \varepsilon < |\nabla \times \mathbf{H}| < 1 + \varepsilon\}$ plays as the intermediate zone and the limit of the zone as $\varepsilon \to 0$ becomes the interface bewteen the normal and superconductor regions.

Unlike the scalar p-Laplacian equation there is not much work being done about the degenerate system (1.1) as well as its limit problem. Recently, the regularity of weak solutions to a linear system analogy to (1.1) was studied in [17]. In this paper we shall first prove the existence of a weak solution to the system (1.1)-(1.3). Then we study the limit solution as $p \to \infty$ and prove that the limit solution solves a variational inequality ([14]). Many techniques are adopted from the scalar p-Laplacian case ([8]). To study the regularity and understand the profile of the limit solution, we consider the magnetic field $\mathbf{H}$ to be plane waves. Then it is rather *surprising* to see that the current density satisfies a porous medium equation. The limit problem for (1.1)-(1.3) becomes the *mesa problem* ( See *Caffarelli-Friedman* [5]) for the density equation. This allows us to analyze the structure of the solution to the limit problem.

The paper is organized as follows. In § 2, by using standard variational method we show that there exists a unique weak solution to the system (1.1) -(1.3). Then we prove that the current density satisfies a porous-medium type equation when the magnetic field is assumed to be plane waves. For this special case, further regularity of the weak solution is obtained. In § 3, we prove that the solution of (1.1)-(1.3) has a unique limit as $p$ goes to infinity. Moreover, the limit solution solves a variational inequality. In § 4, we study the mesa problem with an inhomogeneous term. Similar results to the paper [5] are obtained. In § 5, we use the results obtained in §4 to study the structure of the limit solution of (1.1)-(1.3).

## 2. Existence, Uniqueness and Regularity for fixed $p > 2$

Let $p > 2$. Define
$$B^d(R^3) = \left\{ \mathbf{G}(x) \in W^{1,p}(R^3) : \nabla \cdot \mathbf{G} = 0, \ a.e.x \in R^3 \right\}.$$

We shall assume the following conditions on $\mathbf{H}_0$ and $\mathbf{F}$.

H(2.1) Assume that $\mathbf{H}_0 \in B^d(R^3), \mathbf{F} \in L^2(0, T; B^d(R^3))$.

**Definition 2.1**: *A vector field* $\mathbf{H} \in L^2(0,T; B^d(R^3))$ *is said to be a weak solution of the problem (1.1)-(1.3), if the following integral identity holds:*

$$\int_0^T \int_{R^3} \left[ -\mathbf{H} \cdot \Phi_t + |\nabla \times \mathbf{H}|^{p-2}(\nabla \times \mathbf{H}) \cdot (\nabla \times \Phi) \right] dxdt$$
$$= \int_{R^3} [\mathbf{H}_0(x) \cdot \Phi] \, dx \tag{2.1}$$

*for all* $\Phi(x,t) \in H^1(0,T; B^d(R^3))$ *with* $\Phi(x,T) = 0$ *on* $R^3$.

First of all, we derive some elementary energy estimates.

**Lemma 2.1**: *Under the assumption H(2.1) there exist constants* $C_1$ *and* $C_2$ *such that*

$$\sup_{[0,T]} \int_{R^3} |\mathbf{H}|^2 dx + \int \int_{Q_T} |\nabla \times \mathbf{H}|^p dxdt \le C_1, \tag{2.2}$$

$$\int \int_{Q_T} |\mathbf{H}|^2 dxdt + \frac{1}{p} \sup_{[0,T]} \int_{R^3} |\nabla \times \mathbf{H}|^p dx \le C_2. \tag{2.3}$$

*where* $C_1$ *and* $C_2$ *depend only on known data and the upper bound of* $T$.

**Proof**: Note that for any vector fields $\mathbf{A}, \mathbf{B} \in B^d$ the following identity holds:

$$\int_{R^3} \mathbf{A} \cdot (\nabla \times \mathbf{B}) dx = \int_{R^3} \mathbf{B} \cdot (\nabla \times \mathbf{A}) dx.$$

Taking inner product to the system (1.1) by $\mathbf{H}$ and then using the above identity, we obtain

$$\sup_{[0,T]} \int_{R^3} |\mathbf{H}|^2 dx + \int_0^T \int_{R^3} |\nabla \times \mathbf{H}|^p dxdt$$
$$\le \int_{R^3} |\mathbf{H}_0|^2 dx + \int_0^T \int_{R^3} \mathbf{F} \cdot \mathbf{H} dxdt.$$

By first using Cauchy's inequality and then using Gronwall's inequality, we have

$$\sup_{[0,T]} \int_{R^3} |\mathbf{H}|^2 dx + \int_0^T \int_{R^3} |\nabla \times \mathbf{H}|^p dxdt$$
$$\le \int_{R^3} |\mathbf{H}_0|^2 dx + C \int_0^T \int_{R^3} |\mathbf{F}|^2 dxdt.$$

To obtain the second estimate, we take the inner product by $\mathbf{H}_t$ to (1.1). Similar to the first estimate one can easily derive the second estimate.

$$\text{Q.E. D.}$$

**Theorem 2.2**: *Under the assumption H(2.1) the problem (1.1)-(1.3) has a unique weak solution. Moreover,*

$$\mathbf{H}_t \in L^2(Q_T), \mathbf{H} \in L^\infty(0,T; B^d(R^3)).$$

**Proof**: The proof is based on the standard variational method. For every $\mathbf{V} \in L^2(0, T; R^3)$, define

$$I_p[\mathbf{V}] = \left\{ \begin{array}{l} \frac{1}{p} \int_{R^3} |\nabla \times \mathbf{V}|^p dx, \text{if } \mathbf{V} \in B^d, \\ +\infty, \text{otherwise,} \end{array} \right.$$

for $a.e.$ $t \in (0, T]$. The system (1.1) can be reformulated by the following variational inequality:

$$\mathbf{F} - \mathbf{H}_t \in \partial I_p[\mathbf{H}], x \in R^3, \text{for } a.e. t \in [0, T], \tag{2.4}$$

$$\mathbf{H}(x, 0) = \mathbf{H}_0, \qquad x \in R^3, \tag{2.5}$$

where $\partial I_p[\mathbf{H}]$ is the subdifferential of $I_p[\mathbf{V}]$ at $\mathbf{V}$ (see [16]).

It is clear that the operator $I_p[\mathbf{V}]$ is coercive and strictly monotonic increasing. Moreover, The operator $I_p$ is mapping from $L^2(0, T; W^{1,p}(R^3))$ to $L^2(0, T; W^{-1,p'}(R^3))$, where $p' = \frac{p}{p-1}$. It follows ([16]) that the evolution problem has a unique weak solution $\mathbf{H} \in L^2(0, T; B^d)$. Moreover, by Lemma 2.1 the weak solution $\mathbf{H}_t \in L^2(Q_T)$ and $\nabla \times \mathbf{H} \in L^p(R^3)$ for a.e. $t \in [0, T]$.

$$\text{Q.E.D.}$$

Further regularity of the weak solution seems challenging since there are no local estimates such as those in [6]. However, we can obtain more regularity when the weak solution $\mathbf{H}$ is assumed to be plane waves.

From now on we shall assume that $\mathbf{H}$ depends only on $x = (x_1, x_2)$ and the component in z-direction is zero, i.e., $\mathbf{H}(x, t) = \{h_1(x, t), h_2(x, t), 0\}$. In this case $Q_T = R^2 \times (0, T]$.

To obtain more regularity of the weak solution, we need more regularity for the known data.

H(2.2): Assume that $\mathbf{H}_0 \in C^{2+\alpha}(R^2)$ and $\mathbf{F} \in C^{2+\alpha}(Q_T)$ for some $\alpha \in (0, 1)$.

**Theorem 2.3**: *Under the assumptions H(2.1)-H(2.2), the weak solution of the problem (1.1)-(1.3) belongs to $L^\infty(Q_T) \bigcap C^{1+\alpha, \beta}(Q_T)$ for some $\alpha, \beta \in (0, 1)$. Moreover, the solution has compact support for each $t \in [0, T]$, if $\mathbf{F}$ and $\mathbf{H}_0$ have compact support.*

**Proof**: Recall that $\mathbf{H} = \{h_1(x, t), h_2(x, t), 0\}$. Then it is clear that

$$\nabla \times \mathbf{H} = (h_{2x_1} - h_{1x_2})\mathbf{k},$$

where $\mathbf{k}$ is the unit vector in z-direction. It follows that

$$\nabla \times \left( |\nabla \times \mathbf{H}|^{p-2} \nabla \times \mathbf{H} \right)$$
$$= \left\{ \left[ |h_{2x_1} - h_{1x_2}|^{p-2}(h_{2x_1} - h_{1x_2}) \right]_{x_2}, - \left[ |h_{2x_1} - h_{1x_2}|^{p-2}(h_{2x_1} - h_{1x_2}) \right]_{x_1}, 0 \right\}$$

The system (1.1) is equivalent to the following system:

$$h_{1t} + \frac{\partial}{\partial x_2}\left[|h_{2x_1} - h_{1x_2}|^{p-2}(h_{2x_1} - h_{1x_2})\right] = f_1(x_1, x_2, t), \qquad (2.7)$$

$$h_{2t} - \frac{\partial}{\partial x_1}\left[|h_{2x_1} - h_{1x_2}|^{p-2}(h_{2x_1} - h_{1x_2})\right] = f_2(x_1, x_2, t). \qquad (2.8)$$

From Theorem 2.2, we know that $h_1(-\infty, x_2, t) = 0$ for all $x_2 \in R^1$ and $t \geq 0$. Then from the equation (1.2), we see

$$h_1(x_1, x_2, t) = -\int_{-\infty}^{x_1} h_{2x_2}(\xi, x_2)d\xi = -U_{2x_2}(x_1, x_2, t),$$

where

$$U_2(x_1, x_2, t) = \int_{-\infty}^{x_1} h_2(\xi, x_2, t)d\xi.$$

It follows that

$$h_{1x_2} = -U_{2x_2x_2}, \text{and } h_{2x_1} = U_{2x_1x_1}, (x_1, x_2, t) \in Q_T.$$

Consequently, the equation (2.7) becomes

$$\frac{\partial}{\partial x_2}\left[-U_{2t} + |\Delta U_2|^{p-2}\Delta U_2\right] = f_1(x_1, x_2, t).$$

Integrating over $(-\infty, x_2)$ yields

$$U_{2t} - |\Delta U_2|^{p-2}\Delta U_2 = f_2^*(x, t), \qquad (2.9)$$

where

$$f_2^*(x, t) = -\int_{-\infty}^{x_2} f_1(x_1, \xi, t)d\xi.$$

Similarly, we can eliminate $h_2(x, t)$ to derive the following equation from (2.8):

$$U_{1t} - |\Delta U_1|^{p-2}\Delta U_1 = f_1^*(x, y, t), \qquad (2.10)$$

where

$$U_1(x, t) = \int_{-\infty}^{x_1} h_1(x_1, \xi, t)d\xi,$$

$$f_1^*(x, t) = -\int_{-\infty}^{x_1} f_2(\xi, x_2, t)d\xi.$$

Define

$$u_i(x, t) = \Delta U_i(x, t), g_i(x, t) = \Delta f_i^*(x, t), i = 1, 2.$$

Then $u_1$ and $u_2$ will satisfy the following system:

$$u_{it} - \Delta \psi(u_i) = g_i(x, t), \qquad (x, t) \in Q_T, \qquad (2.11)$$

$$u_i(x, y, 0) = u_{0i}(x), \qquad x \in R^2, \qquad i = 1, 2, \qquad (2.12)$$

where

$$u_{01}(x) = \Delta \int_{-\infty}^{x_2} h_{01}(x_1, \xi)d\xi, u_{02}(x) = \Delta \int_{-\infty}^{x_1} h_{02}(\xi, x_2)d\xi,$$

$$g_1(x, t) = -\triangle \int_{-\infty}^{x_1} f_2(\xi, x_2, t)d\xi, g_2(x, t) = -\triangle \int_{-\infty}^{x_2} f_1(x_1, \xi, t)d\xi,$$

while the function $\psi$ is defined by

$$\psi(s) = |s|^{p-2}s.$$

By a result of [7] (also see [10]), there exists a unique solution

$$u_i(x,t) \in L^2(Q_T) \bigcap L^\infty(Q_T).$$

Moreover, the solution is Hölder continuous in $Q_T$. Furthermore, the $supp\, u_i$ is compact for each $t \in [0,T]$, if $g_i$ and $u_{0i}$ have compact support, $i = 1,2$.

Now for each fixed $t \in (0,T]$, we solve the following elliptic problem:

$$\Delta U_i = u_i, \qquad x \in R^2, \tag{2.13}$$

$$U_i(x,t) = 0, \qquad |x| \to +\infty. \tag{2.14}$$

By applying the standard elliptic theory [11], we know that the problem (2.13)-(2.14) admits a unique solution $U_i(x,t) \in C^{2+\alpha,0}(R^2)$. Now we claim that $U_i(x,t)$ is also Hölder continuous with respect to $t$. Indeed, for small $\triangle t$ we define

$$U_i^* = \frac{U_i(x, t + \triangle t) - U_i(x,t)}{|\triangle t|^{\alpha/2}}.$$

Then $U_i^*$ solves the following problem:

$$\Delta U_i^* = \frac{u_i(x, t + \triangle t) - u_i(x,t)}{|\triangle t|^{\frac{\alpha}{2}}}, x \in R^2, \tag{2.15}$$

$$U_i^*(x,t) = 0, \qquad |x| \to +\infty. \tag{2.16}$$

As $u_i(x,t)$ is Hölder continuous with respect to $t$, we apply $L^p$-theory for elliptic equations to obtain:

$$\|U_i^*\|_{W_p^2(B_R(0))} \le C,$$

where $B_R(0) = \{x \in R^2 : |x| < R\}$ and the constant $C$ depends only on $p$, Hölder norm of $u_i$ and $R$, but independent of $\triangle t$.

By Sobolev's embedding, we know that for every $R > 0$ there exist constants $C(R)$ and $\alpha \in (0,1)$ such that

$$\|U_i^*\|_{C^{1+\alpha}(B_R(0))} \le C(R).$$

A compactness argument yields that $U_i(x,t)$ and $U_{ix_j}$ are Hölder continuous with respect to $t$ for each $i$ and $j$, where $i = 1,2$ and $j = 1,2$. It follows that $h_1(x,t) = \frac{\partial}{\partial x_1}U_1(x,t)$ and $h_2 = \frac{\partial}{\partial x_2}U_2(x,t)$ are Hölder continuous with respect to $t$. Finally, an interpolation lemma ([13]) implies that $h_i$ is Hölder continuous with respect to $t$ with Hölder exponent $\beta$ for some $\beta \in (0,1)$.

<div align="right">Q.E.D.</div>

## 3. THE LIMIT SOLUTION AS $p \to \infty$

In this section we shall show that the solution of (1.1)-(1.3) has a limit as $p \to \infty$, which solves a variational inequality if the initial current density is less than the Bean critical value.

H(3.1): Assume that the initial field $\mathbf{H}_0$ satisfies

$$||\nabla \times \mathbf{H}_0||_{L^\infty(R^3)} \leq 1.$$

**Lemma 3.1**: Under the assumptions H(2.1) and H(3.1), the energy estimates (2.2)-(2.3) hold, where the constants $C_1$ and $C_2$ are independent of $p$.

The proof is the same as for Lemma 2.1 except that the assumption H(3.1) will ensure that $C_1$ and $C_2$ are independent of $p$.

We shall denote by $\mathbf{H}^{(p)}$ the solution of the system (1.1)-(1.3).

From Lemma 3.1, we know that every sequence $\mathbf{H}^{(p')}$ with $p' \to +\infty$ has a subsequence $\mathbf{H}^{(p'')}$ which converges to a limit, denoted by $\mathbf{H}^{(\infty)}$, weakly in $*-L^\infty(\Omega \times (0, T])$ for any bounded domain $\Omega \subset R^3$. Moreover,

$$\nabla \times \mathbf{H}^{(p'')} \to \nabla \times \mathbf{H}^{(\infty)},$$
$$\mathbf{H}_t^{(p'')} \to \mathbf{H}_t^{(\infty)}$$

weakly in $L^2(\Omega \times (0, T))$. Furthermore, since $\nabla \cdot \mathbf{H} = 0$, $\mathbf{H}^{(p)} \in L^2(0, T; B^d)$ and $\mathbf{H}_t \in L^2(Q_T)$, it follows that

$$\mathbf{H}^{(p'')} \to \mathbf{H}^{(\infty)} \qquad a.e. \text{ in } Q_T.$$

**Lemma 3.2**: The limit $\mathbf{H}^{(\infty)}$ satisfies the following estimate:

$$ess \sup_{Q_T} |\nabla \times \mathbf{H}^{(\infty)}| \leq 1.$$

**Proof**: From the proof of Lemma 2.1, we see

$$\int_0^T \int_{R^3} |\nabla \times \mathbf{H}|^p dx dt \leq Cp,$$

where $C$ is a constant which is independent of $p$.

Similar to [8], for every small $\delta > 0$ we define

$$A_\delta = \left\{ (x, t) \in Q_T : |\nabla \times \mathbf{H}^{(p)}| \geq 1 + \delta \right\}.$$

Then,

$$(1 + \delta)|A_\delta| \leq \int \int_{A_\delta} |\nabla \times \mathbf{H}^{(p)}| dx dt$$
$$\leq \lim_{p_i \to \infty} inf \int \int_{A_\delta} |\nabla \times \mathbf{H}^{p_i}| dx dt$$
$$\leq \lim_{p_i \to \infty} inf \left( \int \int_{A_\delta} |\nabla \times \mathbf{H}^{(p_i)}|^{p_i} dx dt \right)^{\frac{1}{p_i}} |A_\delta|^{1 - \frac{1}{p_i}}$$
$$\leq |A_\delta|.$$

It follows that

$$|A_\delta| = 0.$$

The desired result follows since $\delta$ is arbitrary.

Q.E.D.

Before stating the main result in this section, we define another functional. Let

$$K = \{\mathbf{V} \in L^2(Q_T); \nabla \cdot \mathbf{V} = 0\}.$$

For every $\mathbf{V} \in K$, define $I_p[V]$ as in Section 2 for *a.e.* $t \in (0, T]$. Similarly, we define

$$I_\infty[\mathbf{V}] = \left\{ \begin{array}{l} 0, \text{ if } \mathbf{V} \in K, |\nabla \times \mathbf{V}| \leq 1, \\ +\infty, \text{otherwise.} \end{array} \right.$$

**Theorem 3.3**: Under the assumptions H(2.1) and H(3.1) the limit function $\mathbf{H}^{(\infty)}$ is unique and satisfies the following variational problem:

$$\mathbf{F} - \mathbf{H}_t \in \partial I_\infty[\mathbf{H}], \qquad \text{for } a.e.\, t > 0, \tag{3.1}$$

$$\mathbf{H}(x, 0) = \mathbf{H}_0(x), \qquad x \in R^3, \tag{3.2}$$

that is, for a.e. $t \in [0, T]$,

$$I_\infty[\mathbf{V}] - I_\infty[\mathbf{H}^{(\infty)}] \tag{3.2}$$

$$\geq \int_{R^3} (\mathbf{F} - \mathbf{H}_t^{(\infty)}) \cdot (\mathbf{V} - \mathbf{H}^{(\infty)}) dx \qquad a.e.t \geq 0, \tag{3.3}$$

for all $\mathbf{V} \in K$ with $\mathbf{V}_t \in L^2(Q_T)$.

**Proof**: Since the limit is unique we may assume the whole sequence $\mathbf{H}^{(p)}$ converges to $\mathbf{H}^{(\infty)}$. For simplicity, we denote by $\mathbf{H}$ the limit solution. Note that the system (1.1) is equivalent to the following variational form:

$$\mathbf{F} - \mathbf{H}_t \in \partial I_p[\mathbf{H}]. \tag{3.4}$$

By the definition of $I_\infty[\mathbf{V}]$, we only need to verify the inequality for all $\mathbf{V} \in L^2(Q; R^2)$ with $|\nabla \times \mathbf{V}| \leq 1$ a.e. $t \in [0, T]$. On the other hand, Lemma 3.2 implies that $I_\infty[\mathbf{H}] = 0$ for a.e. $t \in [0, T]$. Thus, the inequality (3.3) is equivalent to

$$\int_{R^3} (\mathbf{F} - \mathbf{H}_t) \cdot (\mathbf{V} - \mathbf{H}) \, dx \leq 0.$$

As $\mathbf{H}^{(p)}$ solves the variational problem (2.4)-(2.5), we see

$$\begin{aligned} I_p[\mathbf{V}] &\geq I_p[\mathbf{H}^{(p)}] + \int_{R^2} \left(\mathbf{F} - \mathbf{H}_t^{(p)}\right) \cdot \left(\mathbf{V} - \mathbf{H}^{(p)}\right) dx \\ &\geq \int_{R^3} \left(\mathbf{F} - \mathbf{H}_t^{(p)}\right) \cdot (\mathbf{V} - \mathbf{H}^p) \, dx. \end{aligned} \tag{3.5}$$

By assumptions on $\mathbf{V}$, we have

$$\lim_{p \to \infty} I_p[\mathbf{V}] = 0.$$

On the other hand, since $\mathbf{H}^{(p)} \to \mathbf{H}$ a.e and $\mathbf{H} \in L^2(Q_T)$, it follows that

$$\int_{R^3} \left(\mathbf{F} - \mathbf{H}_t^{(p)}\right) \cdot \left(\mathbf{V} - \mathbf{H}^{(p)}\right) dx \to \int_{R^3} (\mathbf{F} - \mathbf{H}_t) \cdot (\mathbf{V} - \mathbf{H}) \, dx.$$

After taking limit in the inequality (3.5), we obtain the desired inequality (3.3). To verify the initial condition, we note that

$$\int_{R^3} |\mathbf{H}(x, t) - \mathbf{H}_0(x)|^2 dx \leq C\sqrt{t} \to 0$$

as $t \to 0$, since $||\mathbf{H}_t||_{L^2(Q_T)} \leq C$.

Now we prove the uniqueness. Suppose there are two solutions $\mathbf{H}$ and $\mathbf{H}^*$ to the limit problem. Then by Lemma 3.2 we know that

$$|\nabla \times \mathbf{H}| \leq 1, |\nabla \times \mathbf{H}^*| \leq 1.$$

We choose $\mathbf{H}$ and $\mathbf{H}^*$ as test functions in (3.3), respectively, to obtain

$$\int_{R^3}(\mathbf{F} - \mathbf{H}_t) \cdot (\mathbf{V} - \mathbf{H})dx \leq 0,$$

$$\int_{R^3}(\mathbf{F} - \mathbf{H}^*_t) \cdot (\mathbf{V} - \mathbf{H}^*)dx \leq 0,$$

Consequently, one obtains

$$\int_{R^3}(\mathbf{H}^* - \mathbf{H})_t \cdot (\mathbf{H} - \mathbf{H}^*)dx \leq 0 \qquad \text{a.e. } t \geq 0.$$

It follows that

$$\frac{d}{dt}||\mathbf{H} - \mathbf{H}^*||_{L^2(Q_T)} \leq 0,$$

which implies the uniqueness.

<div align="right">Q.E.D.</div>

**Remark 3.1**: The solution obtained in [14] is essentially the same as the limit solution in Theorem 3.3 if the space dimension is equal to 2.

Now we investigate further regularity of the limit solution when $\mathbf{H}$ is a plane wave.

**Theorem 3.4**: *If the space dimension is equal to 2, then the limit solution $\mathbf{H}^{(\infty)}$ is globally bounded and Hölder continuous in $Q_T$.*

**Proof**: When $\mathbf{H}^{(p)} = \{h_1, h_2, 0\}$, we can study the regularity for $u_i^{(p)}$. Recall that for each $t \in [0, T]$,

$$\Delta U_i^{(p)} = u_i^{(p)}(x, t), \qquad x \in R^2, \tag{3.6}$$

$$U_i^{(p)}(x, t) = 0, \qquad |x| \to \infty, \tag{3.7}$$

Since $||u_i^{(p)}||_{L^\infty(Q_T)} \leq 1$,, the elliptic theory implies that for any $q > 1$

$$||U_i^{(p)}||_{W^{2,q}(B_R(0))} \leq C,$$

where $C$ depends only on $R$ and $q$, but not on $p$. It follows by Sobolev's embedding and a compactness argument that for any $\alpha \in (0, 1)$ $U_i^{(\infty)}(x, t) \in C^{1+\beta,0}(Q_T)$ for any $\beta \in (0, 1)$ and each $t \in [0, T]$. It follows from the definition that $\mathbf{H}^{(\infty)} \in C^{\alpha,0}(Q_T)$.

Next we derive the regularity of $U_i^{(\infty)}(x, t)$ with respect to $t$. Recall that $U_i^{(p)}$ satisfies Eq.(3.6). Note that

$$||u_{it}||_{L^1(Q_T)} \leq C,$$

where $C$ is independent of $p$. It follows that for any $q \in (1, 2)$

$$||U_{it}^{(p)}||_{W^{1,q}(Q_T)} \leq C,$$

where $C$ is independent of $p$. As $\mathbf{U}^{(\infty)} \in C^{1+\beta,0}(Q_T)$, it follows by the interpolation result ([13]) that the limit function $U_i^{(\infty)}$ and $U_{ix_l}^{(\infty)}$ are Hölder continuous with respect to $t$.

<div align="right">Q.E.D.</div>

### 4. The Mesa Problem with an inhomogeneous Term

As we have seen in section 2, the limit problem for the system (1.1)-(1.3) becomes the mesa problem for the current density if the space dimension is equal to 2. Throughout this section we shall study the mesa problem with an inhomogeneous term. By employing similar techniques to [5], we derive the profile of the limit solution to the mesa problem. The result will be used to understand the initial collapsing and afterwards evolution process for the current density of the limit solution $\mathbf{H}^{(\infty)}$ in the case of two space dimensions (see Section 5 below).

Consider the mesa problem with an inhomogeneous term.

$$u_t - \Delta u^m = g(x,t), \qquad (x,t) \in R^n \times (0,\infty), \qquad (4.1)$$

$$u(x,0) = f(x), \qquad x \in R^n, \qquad (4.2)$$

H(4.1). Assume that $f(x) \in L^1(R^n) \bigcap L^\infty(R^n)$ and

$$g(x,t) \in L^1(R^n \times (0,\infty)) \bigcap L^\infty(R^n \times (0,\infty)).$$

Moreover, $f(x) \geq 0$ and $g(x,t) \geq 0$ on $Q_T = R^2 \times (0,T]$.

**Lemma 4.1**: *Let*

$$M = ||f||_{L^\infty(R^n)} + T||g||_{L^\infty(Q_T)}.$$

*Then*

$$||u^{(m)}||_{L^\infty(Q_T)} \leq M,$$

*Moreover, if $M < 1$, then $u^{(m)}$ has a unique limit $u^{(\infty)}(x,t)$ as $m \to \infty$ and*

$$u^{(\infty)}(x,t) = f(x) + \int_0^t g(x,\tau)d\tau, (x,t) \in Q_T.$$

**Proof**: As a first step, we may assume that $f(x)$ and $g(x,t)$ have compact support. Then $u^{(m)}(x,t)$ has compact support for each $t \in [0,T]$ and any $m > 1$. Let $R$ be sufficiently large such that $u^{(m)}(x,t) = 0$ on $\partial B_R(0) \times [0,T]$. We may also assume that $u^{(m)}$ is smooth. Define the operator $L$ as follows:

$$L[u] = mu^{m-1}\Delta u + m(m-1)u^{m-2}|\nabla u|^2 - u_t,$$

Let

$$v(x,t) = ||f||_{L^\infty(R^n)} + t||g||_{L^\infty(Q_T)}.$$

Then

$$L[u^{(m)} - v] = g(x,t) + ||g||_{L^\infty(Q_T)} \geq 0.$$

On the parabolic boundary of $B_R(0) \times [0,T]$,

$$u^{(m)} - v \leq 0.$$

By the comparison principle, we can see that

$$u^{(m)}(x,t) \leq v(x,t) \leq M.$$

The lower bound of $u^{(m)}$ can be derived similarly. In general, we use the standard approximation (see page 715 of [5], for example) to derive the desired estimate.

To prove the second conclusion, we note that

$$\int_{R^n} u^{(m)}(x,t)dx - \int_{R^n} f(x)dx = \int_{R^n} \int_0^t g(x,\tau)d\tau dx.$$

It follows that

$$||u^{(m)}||_{L^1(R^n)} \leq ||f||_{L^1(R^2)} + ||g||_{L^1(Q_T)}.$$

For any fixed smooth function $\phi(x)$, we have

$$\int_{R^n} \Delta(u^{(m)})^m \phi(x)dx = \int_{R^n} (u^{(m)})^m \Delta\phi(x)dx$$

$$\leq CM^{m-1}||u^{(m)}||_{L^1(R^n)} \to 0 \qquad \text{as } m \to \infty.$$

It follows that

$$\int_{R^n} u^{(\infty)}(x,t)\phi(x)dx = \int_{R^n} f(x)\phi(x)dx + \int_{R^n} \left[\int_0^t g(x,\tau)d\tau\right]\phi(x)dx.$$

Thus,

$$u^{(\infty)}(x,t) = f(x) + \int_0^t g(x,\tau)d\tau$$

for a.e. $x \in R^n, t \geq 0$.

<div align="right">Q.E.D.</div>

From Lemma 4.1, we know that $u^{(m)}$ is uniformly bounded. Then for any fixed $t > 0$ every sequence $u^{(m')}(x,t)$ with $m' \to \infty$ has a subsequece $u^{(m'')}$ with $m'' \to \infty$ such that

$$u^{(m'')} \to u^{(\infty)}(x,t) \text{ weakly in } *-L^\infty(\Omega).$$

In general $u^{(\infty)}(x,t)$ may not be unique. However, we show that there is a unique limit for $u^{(\infty)}(x,t)$ as $t \to 0+$.

**Lemma 4.2**: Under the assumption H(4.1) the following estimate holds:

$$||u_t||_{L^1(R^n)} \leq \frac{1}{t}||g||_{L^1(R^n)} + \left[\frac{1}{t(m-1)}||u||_{L^1(R^n)} + ||f||_{L^1(R^n)}\right].$$

**Proof**: Let $\hat{u}(x,t) = \lambda^{\frac{1}{m-1}} u(x,\lambda t)$. Then it is easy to see that $\hat{u}$ solves the following problem:

$$\hat{u}_t - \Delta\hat{u}^m = \lambda^{\frac{m}{m-1}} g(x,t), (x,t) \in R^n \times (0,\infty),$$

$$\hat{u}(x,0) = \lambda^{\frac{1}{m-1}} f(x), \qquad x \in R^n$$

By $L^1$-stability for the porous medium equation ([4]) we know that

$$||u - \hat{u}||_{L^1(R^n)} \leq C[(|\lambda^{\frac{m}{m-1}} - 1|||g||_{L^1(R^n)} + |\lambda^{\frac{1}{m-1}-1}|||f||_{L^1(R^n)}],$$

where $C$ depends only on $n$, but not on $m$. By taking limit as $\lambda \to 1+$, we obtain

$$||\frac{1}{m-1}u + tu_t||_{L^1(R^n)}$$

$$\leq \frac{m}{m-1}||g||_{L^1(R^n)} + \frac{C}{m-1}||f||_{L^1(R^n)},$$

which yields the desired estimate.

**Lemma 4.3**: The limit solution satisfies

$$0 \le u^{(\infty)} \le 1 \qquad \text{in } Q_T.$$

**Proof**: For any $\delta > 0$, we multiply the equation (4.1) by $\hat{u} = [u - (1+\delta)]^+$ to obtain that

$$\int_{R^n} |\hat{u}|^2 dx + \int_0^T \int_{R^n} m u^{m-1} |\nabla \hat{u}|^2 dx dt$$

$$\le \int_{R^n} |\hat{u}(x,0)|^2 dx + \int_0^T \int_{R^n} |g|^2 dx dt \le C,$$

where $C$ is independent of $m$.

It follows that

$$\int_0^T \int_{R^n} |\nabla \hat{u}|^2 dx dt \le \frac{C}{m}.$$

Fatou's lemma implies

$$|\{(x,t) : u^{(\infty)} \ge 1 + \delta\}| = 0,$$

which yields the desired result since $\delta$ is arbitrary.

<div align="right">Q.E.D.</div>

From Lemma 4.3, we see that there must exist an initial collapsing instantly if $||f||_{L^\infty(R^n)} > 1$. The external force should play no role in this collapsing process. The basic strategy is to separate the initial collapsing and the evolution processes. H(4.2): Let $L = ||f||_{L^\infty(R^n)} > 1$.

**Theorem 4.4**: *Under the assumptions H(4.1)-(4.2), $u^{(m)}(x,t)$ has a unique limit as $m \to \infty$ and $t \to 0+$. Moreover,*

$$u^{(\infty)}(x, 0+) = v^{(\infty)}(x), \qquad x \in R^n,$$

*where $v^{(\infty)}(x)$ is the limit function of the solution $v^{(m)}$ to (4.1)-(4.2) with $g(x,t) \equiv 0$.*
**Proof**: Let $v^{(m)}$ be the solution of (4.1)-(4.2) with $g(x,t) = 0$. Let $\phi(x)$ be any bounded function with compact support. Let $(m', t_{m'})$ and $(m'', t_{m''})$ be any two sequences pairs which converge to $(\infty, 0+)$, respectively, as $m \to \infty$. For any bounded function $\phi$

$$\left| \int_{R^n} \phi \left[ u^{(m')}(x, t_{m'}) - v^{(m'')}(x, t_{m''}) \right] dx \right|$$

$$\le ||\phi||_{L^\infty(R^n)} \int_{R^n} [\int_0^{t_{m'}} |u_t^{(m')}| d\tau + \int_0^{t_{m''}} |v_t^{(m'')}| d\tau] dx$$

$$\le C ||\phi||_{L^\infty(R^n)} \left[ \int_0^{max\{t_{m'}, t_{m''}\}} \int_{R^n} |g| dx d\tau + \frac{C}{m'-1} + \frac{C}{m''-1} \right]$$

$$\to 0,$$

as $m', m'' \to \infty$. Let the limit function be denoted by $u^{(\infty)}(x, 0+)$.

From [5], we know that $v^{(m)}(x,t)$ converges to a unique limit $v^{(\infty)}(x)$ as $m \to \infty$. Particularly, we see that

$$v^{(m'')}(x, t_{m''}) \to v^{(\infty)}(x, 0+) \equiv v^{(\infty)}(x)$$

as $m'' \to \infty$.

It follows that

$$\int_{R^n} \phi u^{(\infty)}(x, 0+) dx = \int_{R^n} \phi v^{(\infty)}(x, 0+) dx.$$

Since $\phi(x)$ is arbitrary, we see that

$$u^{(\infty)}(x, 0+) = v^{(\infty)}(x, 0+) = v^{(\infty)}(x), \qquad a.e. x \in R^n.$$

<div align="right">Q.E.D.</div>

**Remark 4.1**: Physically, Theorem 4.3 means that the collapsing process happens instantly before the force term $g(x, t)$ affects the evolution process.

To see the profile of the limit solution for $t > 0$, we may start with $u^{(\infty)}(x, 0+)$ as an initial value. Therefore, without loss of generality, from now on we assume that

$$L = ||f||_{L^\infty(R^n)} \leq 1.$$

**Lemma 4.5** Assume that $||f||_{L^\infty(R^n)} \leq 1$. Then

$$V_m = \frac{m}{m-1} \left( u^{(m)} \right)^{m-1},$$

is uniformly bounded.

**Proof**: With loss of generality, we may assume that $V_m$ is smooth and $supp V_m \subset B_R(0)$ for all $t \in [0, T]$. A direct calculation shows that $V_m$ satisfies

$$V_{m,t} = (m-1)V_m \Delta V_m + |\nabla V_m|^2 + V_m^\beta g(x, t), x \in B_R, t \in (0, T),$$

$$V_m(x, t) = 0, \qquad (x, t) \in \partial B_R(0) \times [0, T],$$

$$V_m(x, 0) = \frac{m}{m-1} f(x)^{m-1}, x \in R^n,$$

where $\beta = \frac{m-2}{m-1} < 1$. If $V_m$ attains positive maximum $M > 1$, then the coefficient of $V_m$, $\frac{g}{V_m^{1-\beta}}$, has an upper bound which is independent of $m$. The strong maximum principle implies that $V_m$ can not take a positive maximum which is greater than 1 in $B_R(0) \times (0, T]$. It follows that

$$V_m \leq max\{1, \max_{B_R(0)} V_m(x, 0)\} \leq \frac{m}{m-1} \leq 2.$$

<div align="right">Q.E.D.</div>

**Lemma 4. 6**: $u^{(m)}$ has a uniform compact support.

**Proof**: By Lemma 4.5, $V_m$ is uniformly bounded. Now consider the special Barenblatt solution

$$\hat{u} = t^{-\frac{n}{n(m-1)+2}} \left[ \xi_0^2 - \frac{m-1}{2[n(m-1)+2]} |x|^2 t^{-\frac{2}{n(m-1)+2}} \right]_+^{\frac{1}{m-1}},$$

where $(s)_+ = max\{s, 0\}$ and

$$\xi_0 = \left[ \frac{M}{w_n} \left( \frac{2(n(m-1)+2)}{m-1} \right)^{-\frac{n}{2}} \left( \int_0^1 (1-y^2)^{\frac{1}{m-1}} y^{n-1} dy \right)^{-1} \right]^{\frac{m-1}{n(m-1)+2}}.$$

Assume that $suppf, suppg \subset B_R(0)$ for some large $R$ and for every $t \in [0, T]$. By choosing $M$ sufficiently large and using the comparison principle in the domain $(R^n \backslash B_R(0)) \times (0, T]$, we obtain the desired result.

<div align="right">Q.E.D.</div>

H(4.3): Assume that $f(x) \in C^1(R^n)$ is radially symmetric with compact support and $f_r < 0$ in $R^n \backslash \{0\} \bigcap suppf$. Assume that $g(x, t) \in C^{2,1}(R^n \times [0, T])$ is radially symmetric in $x$ and $g_r \leq 0$ in $R^n \backslash \{0\}) \bigcap suppg$ for each $t \in [0, T]$. Moreover,

$$\Delta f^m + g(x, 0) \geq 0.$$

**Lemma 4.7**: Under the assumptions H(4.1)-(4.3), $u_r^{(m)} \leq 0$ and $u_t^{(m)} \geq 0$.
**Proof**: This can be derived by a simple application of the maximum principle.

**Theorem 4.8**: Under the assumptions H(4.1) and H(4.3), the limit solution $u^{(\infty)}$ is uniquely determined and

$$u^{(\infty)}(x, t) = \begin{cases} 1, & \text{in } \{(x, t) : w(x, t) > 0\}, \\ f(x) + \int_0^t g(x, \tau) d\tau, & \text{in } \{(x, t) : w(x, t) = 0\}, \end{cases}$$

where $w(x, t)$ is the unique solution of the variational inequality

$$-\Delta w \geq f(x) + \int_0^t g(x, \tau) d\tau - 1, \tag{4.3}$$

$$w(x, t) \geq 0, \tag{4.4}$$

$$w(x, t)[\Delta w + f(x) + \int_0^t g(x, \tau) d\tau - 1] = 0 \tag{4.5}$$

for a.e. $(x, t) \in R^n \times (0, T]$.
**Proof**: Since the proof closely follows the argument of [5] (Theorem 5.2-5.3), we only give a brief outline here. First of all we recall that $||f||_{L^\infty(R^n)} \leq 1$. Since $f_r \leq 0$, we may further assume that $f(x) = 1$ on a small ball, say, $B_\delta(0)$ for some small $\delta \geq 0$. Since $g(x, t) \geq 0$, by the comparison principle (see [5] Theorem 3.2) we see that $u^{(\infty)}(x, t) = 1$ for a.e. $x \in B_\delta$ and $t \in [0, T]$.

Step 1: For each $t \in [0, T]$, define

$$N_t = \{x : f(x) + \int_0^t g(x, \tau) d\tau = 1\}.$$

By the comparison principle, it is clear that $N_{t_1} \subset N_{t_2}$ if $t_1 < t_2$. We claim that

$$u^{(\infty)}(x, t) = 1, \text{for a.e. all } x \in N_t.$$

Indeed, we replace $f$ and $g$ by $f_\varepsilon = min\{1 - \varepsilon, f\}$ and $g - \varepsilon$, respectively, for small $\varepsilon > 0$. Let $u_\varepsilon^{(m)}$ and $u_\varepsilon^{(\infty)}$ be the solutions of (4.1)-(4.2) corresponding to $f_\varepsilon$ and $g_\varepsilon$, respectively. Then Lemma 4.1 yields

$$u_\varepsilon^{(\infty)}(x, t) = f_\varepsilon(x) + \int_0^t g_\varepsilon(x, \tau) d\tau$$

for a.e. all $x \in R^n$. On the other hand, the comparison principle implies

$$u^{(m)}(x,t) \geq u_\varepsilon^{(m)}(x,t).$$

It follows that

$$u^{(\infty)}(x,t) \geq f_\varepsilon(x) + \int_0^t (g - \varepsilon)d\tau.$$

It follows by the definition of $N_t$ that

$$u^{(\infty)}(x,t) = 1, \qquad \text{as } \varepsilon \to 0,$$

for a.e. $x \in N_t$.

Step 2: Let

$$S_t = \{x : u^{(\infty)}(x,t) < 1\},$$
$$S_t^* = \{x \in S_t : S_t \text{ has Lebesgue density 1 at } x\}.$$

Assume that for any $x_0 \in S_t^*$ and $t_0 \in (0,T]$ with $u^{(\infty)}(x_0,t_0) \leq \theta_0 < 1$. Then for any $\theta \in (\theta_0,1), \delta > 0$, there exists an $m_0$ such that

$$\inf_{x \in Q_\delta(x_0,t_0)} u^{(m')}(x,t) \leq \theta,$$

if $m' > m_0$.

The proof is almost the same as the proof of Theorem 3.3 in [5]. We skip it here.

Step 3: If $x_0 \in S^*$ with $u^{(\infty)}(x_0,t_0) \leq \theta_0 < 1$, then

$$u^{(\infty)}(x,t) = f(x) + \int_0^t g(x,\tau)d\tau$$

in the $(R^n \backslash B_{r_0}(0)) \times [0,T_0]$, where $r_0 = |x_0|$.

This step can be proved by using Step 3 and Lemma 4.7. Indeed, by Step 2 we know that for any $\theta \in (\theta_0,1)$ and any $\delta > 0$ there exists $m_0$ such that

$$\inf_{B_\delta(x_0)} u^{(m')}(x,t_0) < \theta$$

if $m' > m_0$. Suppose that the minimum of $u^{(m')}$ attains at a point $(x^*,t_0)$ in $B_\delta(x_0)$. Since $u_r^{(m)} \leq 0$ and $u_t^{(m)} \geq 0$, $x^*$ must occur on the boundary of $B_\delta(x_0)$ and, if $m' > m_0$, then

$$u^{(m')}(x,t) \leq \theta$$

for all $(x,t) \in (R^n \backslash B_{r_0+\delta}(0)) \times [0,t_0]$, where $r_0 = |x_0|$. It follows by Lemma 4.1 and Lemma 4.7 that

$$u^{(\infty)}(x,t) = f(x) + \int_0^t g(x,\tau)d\tau$$

for all $(x,t) \in (R^n \backslash B_{r_0+\delta}(0)) \times [0,t_0]$. Finally, since $\delta$ is arbitrary, we conclude the desired result.

Step 4:

$$u^{(\infty)}(x,t) = \chi_{N_t} + [f(x) + \int_0^t g(x,\tau)d\tau]\chi_{R^n \backslash N_t}.$$

Indeed, by Step 1 to Step 3, there exists a function $0 \leq \hat{f}(x,t) < 1$ such that

$$u^{(\infty)}(x,t) = \chi_{N_t} + \hat{f}(x,t)\chi_{R^n \backslash N_t}.$$

If at a point $(x_0, t_0)$ with

$$f(x_0) + \int_0^{t_0} g(x, \tau) d\tau < 1,$$

then by Lemma 4.1 and Lemma 4.7,

$$u^{(\infty)}(x, t) = f(x) + \int_0^t g(x, \tau) d\tau$$

for all $x \in R^n \backslash B_{r_0}(0)$. It follows that

$$\hat{f}(x, t) = f(x) + \int_0^t g(x, \tau) d\tau.$$

Step 5:

$$N_t = \{x : w(x, t) > 0\},$$

where $w(x, t)$ is the solution of the variational inequality (4.3)-(4.5).

First of all, by Lemma 4.6 we know that $u^{(\infty)}$ has compact support. Define

$$w(x_0, t) = \int_{R^n} \left( f(x) + \int_0^t g(x, \tau) - u^{(\infty)}(x, t) \right) \Gamma_{x_0} dx,$$

where $\Gamma_{x_0}$ is the fundamental solution with singularity at $x_0$. Note that $w(x_0, t)$ is well defined since $u^{(\infty)}$ has compact support. It is clear that

$$-\Delta w = f(x) + \int_0^t g(x, \tau) d\tau - 1$$

in $N_t$ (since $u^{(\infty)}(x, t) = 1$ on $N_t$).

On the other hand,

$$
\begin{aligned}
w^{(m)}(x_0, t) &= \int_{R^n} [u^{(m)}(x, t) - f(x) - \int_0^t g(x, \tau) d\tau] \Gamma_{x_0} dx \\
&= \int_0^t \int_{R^n} \Delta (u^{(m)})^m \Gamma_{x_0} dx d\tau \\
&= \int_0^t (u^{(m)})^m (x_0, \tau) d\tau.
\end{aligned}
$$

Clearly, $w^{(m)}(x_0, t) \geq 0$ for any $x_0 \in R^n$ and $t \in [0, T]$. By Step 4, we see that if $x_0 \in S_t$, then there exists a small ball $B_{\delta'}(x_0) \subset S_t$. This implies

$$u^{(m')}(x_0, \tau) \leq \theta < 1$$

for all $\tau \in [0, t]$, provided that $m'$ is sufficiently large. Consequently, $w^{(m)}(x_0, t) \to 0$ as $m \to \infty$. Thus,

$$
\begin{aligned}
&w(x, t) \geq 0, \text{ for a.e. } x \in R^n, t \in [0, T], \\
&w(x_0, t) = 0, \qquad \text{if } x_0 \in N_t,
\end{aligned}
$$

i.e. $w(x, t)$ is a solution of the variational inequality (4.3)-(4.5).

Q.E.D.

## 5. The Profile of the Current Density for Limit Solution

Throughout this section we assume that

$$\mathbf{H} = \{h_1, h_2, 0\}.$$

We shall use the results from the previous section to derive the profile of the current density.

H(5.1): Assume that

$$L = ||\nabla \times \mathbf{H}_0||_{L^\infty(R^2)} > 1.$$

H(5.2): Assume that $\mathbf{F}$ and $\mathbf{H}_0$ satisfy the assumptions H(4.1)-H(4.2), where $u_{0i}$ and $g_i$ are defined as in section 2.

First of all, since

$$\nabla \mathbf{H}_0 = h_{01x_1} + h_{02x_2} = 0,$$

we see by the definition that

$$u_{01}(x) = u_{02}(x) = h_{02x_2}(x) - h_{01x_1}(x).$$

Similarly, since $\nabla \times F = 0$ one can find

$$g_1(x, t) = g_2(x, t), \text{ in } Q_T.$$

Define the current densities $J_p$ and $J$ as follows:

$$J_p(x, t) = |\nabla \times \mathbf{H}^{(p)}|, J(x, t) = |\nabla \times \mathbf{H}^{(\infty)}|.$$

Note that $J$ is well-defined since $\nabla \times \mathbf{H}^{(\infty)}$ is well-defined.

**Theorem 5.1**: *Under the assumptions H(5.1)-(5.2) the current density $J(x, 0+) = 1$ if $x \in S_0$ and $J(x, 0+) = |\nabla \times \mathbf{H}_0|$ if $x \in N_0$, where $N_0 = R^2 \backslash S_0$, $S_0$ is the noncoincident set of a variational inequality*

$$-\Delta w \geq g - 1, w \geq 0, (\Delta w + g_i - 1)w = 0.$$

**Proof**: From Theorem 4.3, we know that

$$u_i^{(\infty)}(x, 0+) = v^{(\infty)}(x).$$

On the other hand, we know that

$$u_i^{(p)} = \Delta U^{(p)} = |\nabla \times \mathbf{H}^{(p)}|.$$

It follows

$$J_p(x, t) \to J(x)$$

as $p \to +\infty$. Thus, in the supercounductor region, denoted by $S_0$, the current density

$$J(x, 0+) = v^{(\infty)}(x) = |\nabla \times \mathbf{H}_0|, x \in R^2$$

In the normal conductor region, denoted by $N_0$, the current density

$$J(x, 0+) = v^{(\infty)}(x) = 1.$$

Moreover, $S_0$ is the noncoincident set of a variational inequality (see [5]):

$$-\Delta w \geq u_{0i} - 1, w \geq 0, (\Delta w + u_{0i} - 1)w = 0.$$

Q.E.D.

Next we assume that

$$||\nabla \times \mathbf{H}_0||_{L^\infty(R^2)} \leq 1.$$

H(5.2)': Assume that $\mathbf{H}_0$ and $\mathbf{F}$ satisfy the assumptions H(4.3), where $u_{0i}$ and $g_{0i}$ are defined the same as in Section 2.

**Theorem 5.2**: *Under the conditions H(5.1)-H(5.2)' the current density $J(x,t) = |\nabla \times \mathbf{H}^{(\infty)}|$ has the following structure:*

$$J(x,t) = \begin{cases} 1, \text{ if } x \in N_t, t > 0, \\ f(x) + \int_0^t g(x,\tau)d\tau, \text{ if } x \in S_t, t > 0. \end{cases}$$

*where $N_t = R^2 \backslash S_t$ and $S_t$ is the noncoincident set of the variational inequality*

$$w(x,t) \geq 0, -\Delta w \geq u_{0i} + \int_0^t g_i d\tau - 1,$$

$$w(\Delta w + u_{0i} + \int_0^t g_i d\tau - 1) = 0.$$

**Proof**: From Theorem 4.8, we see that

$$J(x,t) = u_i^{(\infty)}(x,t).$$

The rest of the proof follows the same procedure as that of Theorem 5.1.

Q.E.D.

Next we show the limit solution of (1.1)-(1.3) satisfies a degenerate evolution system.

**Theorem 5.3**: *There exists a nonnegative, bounded and measurable function $a(x,t)$ such that the limit solution $\mathbf{H}^{(\infty)}$ satisfies*

$$\mathbf{H}_t^{(\infty)} - \nabla \times [a(x,t)\nabla \times \mathbf{H}^{(\infty)}] = \mathbf{F}, (x,t) \in Q_T,$$

*in the sense of distribution. Moreover,*

$$supp \, a \subset N_t \qquad for \ a.e. \ every \ t \in (0,T],$$

**Proof**: The proof is similar to [9]. Define

$$\mathbf{A}_p = |\nabla \times \mathbf{H}|^{p-2}\nabla \mathbf{H}.$$

Then, for any $\mathbf{V} \in H(0,T;B^d)$ with $supp\mathbf{V} \in B_R(0)$ for each $t \in (0,T]$, we have

$$\int_0^T \int_{B_R(0)}[-\mathbf{H}^{(p)} \cdot \mathbf{V}_t + \mathbf{A}_p \cdot \nabla\mathbf{V}]dxdt$$

$$= \int_0^T \int_{B_R(0)} \mathbf{F} \cdot \mathbf{V}dxdt.$$

By Lemma 4.6, we know that there exists a subsequence such that

$$\mathbf{A}_{p_k} \to \mathbf{A}, \qquad \text{weakly-* in } L^\infty(R^2 \times (0,T)).$$

It follows that

$$\int_0^T \int_{B_R(0)} [-\mathbf{H}^{(\infty)} \cdot \mathbf{V}_t + \mathbf{A} \cdot \nabla \times V] dx dt$$
$$= \int_0^T \int_{B_R(0)} \mathbf{F} \cdot \mathbf{V} dx dt.$$

On the other hand, we take $\mathbf{V} = \mathbf{H}^{(p)}$ as a test vector to obtain

$$\frac{1}{2} \int_{B_R(0)} |\mathbf{H}^{(p)}|^2 dx + \int_0^T \int_{B_R(0)} |\nabla \mathbf{H}^{(p)}|^p dx dt$$
$$= \int_0^T \int_{B_R(0)} \mathbf{F} \cdot (\mathbf{H}^{(p)}) dx dt \rightarrow \int_0^T \int_{B_R(0)} \mathbf{F} \cdot (\mathbf{H}^{(\infty)}) dx dt.$$

It follows that

$$\int_0^T \int_{B_R(0)} |\nabla \times \mathbf{H}^{(p)}|^p dx dt \rightarrow \int_0^T \int_{B_R(0)} \mathbf{A} \cdot (\nabla \times \mathbf{H}) dx dt$$

as $p \rightarrow \infty$.

By Fatou's lemma, we have

$$\int_0^T \int_{B_R(0)} |\mathbf{A}| dx dt \leq limin f_{p_k \rightarrow \infty} \int_0^T \int_{B_R(0)} |\mathbf{A}_{p_k}| dx dt$$
$$\leq \lim_{p_k \rightarrow \infty} \left( \int_0^T \int_{B_R(0)} |\nabla \times \mathbf{H}^{(p_k)}|^{p_k} dx dt \right)^{1 - \frac{1}{p_k}} [T|B_R(0)|^{\frac{1}{p_k}}$$
$$= \int_0^T \int_{B_R(0)} \mathbf{A} \cdot (\nabla \times \mathbf{H}^{(\infty)} dx dt.$$

It follows that

$$|\mathbf{A}| = \mathbf{A} \cdot (\nabla \times \mathbf{H}^{(\infty)}),$$

since $|\nabla \times \mathbf{H}^{(\infty)}| \leq 1$. Consequently, there exists a nonnegative, bounded function $a(x, t)$ such that

$$\mathbf{A}(x, t) = a(x, t) \nabla \times \mathbf{H}^{(\infty)}.$$

Moreover, if $\nabla \times \mathbf{H}^{(\infty)}$ exists and $|\nabla \times \mathbf{H}^{(\infty)}| < 1$, then $a(x, t) = 0$.

<div align="right">Q.E.D.</div>

## Acknowledgment

The author would like to thank Professor L. C. Evans and Dr. M. Feldman for many helpful discussions during the preparation of this paper.

This work is done at MSRI, University of California, Berkeley, CA 94720. The author would also like to thank Directors and Staff at MSRI for providing excellent research environment.

Department of Mathematics, University of Notre Dame, Notre Dame, IN 46556
*Current address*: Department of Mathematics, University of California, Berkeley, CA 94720